\documentstyle[11pt,nyas]{article}

\input psfig

\def\approxgt{\,\raise2pt \hbox{$>$}\kern-8pt\lower2.pt\hbox{$\sim$}\,}
\def\approxlt{\,\raise2pt \hbox{$<$}\kern-8pt\lower2.pt\hbox{$\sim$}\,}

 \def\ah{{\scriptstyle{1/2}}}
 \def\trh{{\scriptstyle{3/2}}}
 
 \def\twot{{\scriptstyle{{2\over 3}}}}

 \def\ahh{{\scriptscriptstyle 1/2}}
 \def\trhh{{\scriptscriptstyle{3/2}}}

 \def\th{\thinspace}
 \def\ngth{\negthinspace}
 \def\ni{\noindent}
 
 \def\Teff{{$T_{ef\!f} $}}
 \def\Mo{{M$_\odot $}}
 \def\Lo{{L$_\odot $}}

 \def\etal{{et al.~}}
 \def\viz{{viz.~}}
 \def\ul#1{{\underbar {#1}}}

 \def\dotd{\hbox{$.\!\!^{\rm d}$}}

 \long\def\jumpover#1{{}}

\begin{document}

\title{Turbulent Convection in the Classical Variable Stars}
\author{J. Robert Buchler} 
\affil{Physics Department, University of Florida,
    Gainesville, FL 32611}
\author{Zolt\'an Koll\'ath}
\affil{Konkoly Observatory, Budapest, HUNGARY}

\begin{abstract}
\textbf{ We give a status report of convective Cepheid and RR Lyrae model
pulsations.  Some striking successes can be reported, despite the use of a
rather simple treatment of turbulent convection with a 1D time-dependent
diffusion equation for the turbulent energy.  It is now possible to obtain
stable double-mode (beat) pulsations in both Cepheid and RR Lyrae models with
astrophysical parameters, i.e. periods and amplitude ratios, that are in
agreement with observations.  The turbulent convective models, however, have
difficulties giving global agreement with the observations.
In particular, the Magellanic Cloud Cepheids, that have been observed in
connection with the microlensing projects have imposed novel
observational constraints because of the low metallicity of the MCs.  }
\end{abstract}


\section{Introduction}

Cepheid and RR Lyrae variables have played a central role in astrophysics.
This is not only because of interest in the pulsation mechanism itself, and
of the difficulty one had to reconcile them with stellar evolution and stellar
pulsation (Cox 1980), but also, in the case of the Cepheids, because of the
essential role they have played as a local distance indicator in the quest for
determining the Hubble constant. As a result they are the best known and
studied stellar pulsators (for a recent review of stellar pulsations, cf.
Gautschy \& Saio 1995).

Recently, our observational base of Cepheids has been greatly enlarged with the
data from the microlensing projects EROS, MACHO and OGLE (e.g. Ferlet et
al.\ngth 1996).  Not only has the number of well observed Cepheids increased
dramatically through these efforts, we now also have a much broader data base,
particularly because the Magellanic Cloud (MC) Cepheids have metallicities that
range from one half to as low as one quarter solar.

This increased observational knowledge of Cepheids has also brought to light
new problems in the modelling (e.g. Buchler 1998).  For example, {\sl purely
radiative} models give light and radial velocity curves in good agreement with
the observations of the Galactic Cepheids (Moskalik \etal 1992).  But is no
longer true for the Magellanic Cloud Cepheids with their low metallicities.  

The reader may ask here why theorists have bothered to study purely radiative
models (in which heat transport is assumed to be through radiation transfer
only) when we know that convection plays a role in these stars, notably in
diminishing vibrational driving in the colder models and in bringing about the
red edge of the instability strip.  The reason is that radiative models have
been expected to give excellent light curves and radial velocity curves because
convective transport was deemed to be not very efficient nor presumably
important overall.  And, indeed, this expectation was satisfied for Galactic
Cepheid models.  However, despite great efforts, it has not been possible to
model the MC Cepheids with purely radiative models, and it is fair to say that
all possible reasons for the discrepancies have now been exhausted (Buchler
1998).  For a similar conclusion about RR~Lyrae see Kov\'acs \& Kanbur (1997).

We have therefore been led to consider convective transport in our models.  The
physical conditions are such that one expects well developed turbulence: the
Rayleigh number in the stellar envelope is huge and the Prandtl number is tiny.
Possibly plumes play an important role (Rieutord \& Zahn 1995, Zahn, this
Volume).

We are not merely interested in computing a convective flux in a static model,
we also want the linear eigenvalues in order to gain knowledge about the
periods of the models and their stability.  Finally, we want to be able to do
hydrodynamics so as to compute the longterm pulsational behavior of the models,
whether period or multiperiodic (as for the Beat Cepheids, q.v. below).  Since
we are interested in the pulsational problem which is complicated enough in
itself we need a simple, but physically and mathematically robust recipe for
the description of turbulence and convection.

\section{The Equations}

The Cepheids are radial pulsators in which the centrally condensed core is
inert, and only the envelope undergoes pulsational motion.  The motions are
thus governed by 1D hydrodynamics in spherical geometry, supplemented by a
description for the coupling of turbulence and convection with the fluid
dynamics.

\vskip 9pt

 \ul{Hydrodynamic} \ul{Equations}

 \begin{eqnarray}
  {du\over dt} &=& -{1\over\rho}{\partial \over\partial r}
\left(p+\th\th p_t+p_\nu\th\th \right)
   - {G M_r\over r^2} \quad\quad \\
 & & \nonumber \\
   {d\th e\over dt} +p\th  {d\th v\over dt}
  &=& -{1\over\rho r^2} {\partial \over\partial r} \left[ r^2
\left(F_r+\th\th F_c\th\th \right)\right]  -\th \mathbf{{\cal C}}
 \quad\quad 
 \end{eqnarray}

 \vskip 1cm

Convection interacts with the hydrodynamics of the radial motion through the
convective flux $F_c$ term, through a viscous eddy pressure $p_\nu$ and a
turbulent pressure $p_t$, and finally, through an energy coupling term ${\cal
C}$.  The question is how to approximate these quantities in the simplest
physically acceptable way, so as to make it still possible to compute nonlinear
stellar pulsations.

The simplest recipe involves a single, time-dependent diffusion equation for
the turbulent energy $e_t$, of the form

\vskip 25pt

 \ul{Turbulent} \ul{Energy} \ul{Equation}

\vskip 15pt

 \begin{equation}
   {de_t\over dt} +
    \left(p_t+p_\nu\right)\th  {d\th v\over dt}
  = -{1\over\rho r^2} {\partial \over\partial r}\left( r^2 F_t\right)
   +{\cal C} \quad\quad \nonumber\\
 \label{eqs_tc}
\end{equation}

\vskip 9pt

\noindent The coupling term has the general form \begin{equation} {\cal C} =
{\cal S} - \epsilon \end{equation} In the spirit of Kolmogorov the dissipation
rate $\epsilon$ is taken to be \begin{equation} \epsilon = \alpha_d \th \th
e_t^\trhh/\Lambda \label{ML} \end{equation} where $\Lambda = \alpha_\Lambda
H_p$ is the mixing length, proportional to the pressure scale height $H_p= d\th
\ln p/dr = p/(\rho g)$.  We define the turbulent pressure $p_t=\alpha_p \th
\rho \th e_t$ and the viscous eddy pressure $p_\nu = 4/3
\alpha_\nu\th\alpha_\Lambda\th H_p\th \sqrt{e_t}\th r(\partial (u/r)/\partial
r)$.

There is however no unique way of defining $F_c$ and ${\cal S}$ in terms of
 $e_t$ and $Y$, and several variants have been used.  Stellingwerf (1982),
 Kuhfu\ss (1986), Gehmeyr (1992), Gehmeyr \& Winkler (1992), Feuchtinger
 (1998a), Bono \& Stellingwerf (1994), Bono \etal (1997, 1999) and Yecko \etal
 (1998, hereafter YKB) all have used such a 1D equation in the computation of
 nonlinear pulsations, but they have made different possible choices for the
 dependences of the convective flux $F_c$ and for the coupling term ${\cal C}$
 on the turbulent energy $e_t$ and on the dimensionless entropy gradient

 \begin{equation}
 Y = -{H_p\over c_p} \th\th {ds\over dr}
 \end{equation}

\ni With the definitions
 \begin{eqnarray}
 A & = & \alpha_c \th \alpha_\Lambda\th \rho c_p T \\
 B & = & \alpha_s \th \alpha_\Lambda \th
 \sqrt{p \beta T\over \rho} = \alpha_s\alpha_\Lambda \sqrt{{\beta T\over
 \Gamma_1}} \th\th c_s
 \end{eqnarray}
 where $\beta = (\partial \ln v / \partial T)_p$ and $c_s$ is the sound speed,
 we can write the three schemes as

 \begin{eqnarray}
 F_c & = & A \th \th  e_t^\ahh Y  \ \nonumber \\
 {\cal S} & = & \alpha_d
    B^2 \th e_t^\ahh \th Y / \Lambda  \hskip 2cm (GW) \nonumber \\
   & \ & \ \nonumber \\
 F_c & = & A \th\th  e_t^\ahh \th Y \ \nonumber\\
 {\cal S} & = &  \alpha_d
       B \th e_t \th Y^\ahh / \Lambda    \hskip 2cm   (YKB) \nonumber \\
   & \ & \  \nonumber \\
 F_c & = & A/B \th \th e_t\th Y^\ahh \ \nonumber \\
 {\cal S} & = &  \alpha_d B\th e_t\th Y^\ahh / \Lambda
          \hskip 2cm (S) \quad  \\
   & \ & \ \nonumber 
 \label{tc_rec}
 \end{eqnarray}

 The combinations of $A$ [energy/volume] (essentially the internal energy) and
$B$ [velocity] (essentially the sound speed $c_s$) have been chosen so that in
the absence of diffusion all three schemes give the same static equilibrium
model. (Generally they do not give same linear vibrational eigenvalues though).
The model involves seven dimensionless $\alpha$ parameters of order unity.

 All three recipes can be parametrized with a coefficient $N$
such that for
 \begin{eqnarray}
 N > 0:\quad F_c & = & A \th e_t^\ahh \th Y \nonumber \\
 N < 0:\quad F_c & = & A/B \th e_t \th Y^\ahh 
 \end{eqnarray}
 and, with $b\equiv 1/|N|$,
 \begin{eqnarray}
 {\cal S} & = & \alpha_d\alpha_s\th\th B^{2b} \th e_t^{\trhh -b} \th Y^b 
                 \th/\Lambda
 \label{N}
 \end{eqnarray}

\ni Thus we have\th\th
Gehmeyr-Winkler (or Kuhfu\ss) : $N$=1,\th\th
YKB:  $N$=2,\th\th
Stellingwerf: $N$=--2.
(Note that one more unused possibilities exist, \viz $N$=--1).

It is of course interesting to see the influence of the chosen recipe on the
stellar pulsation, which we address in the next section.

At this point we should note that of course more complicated recipes have been
suggested, none of which have been implemented in nonlinear stellar pulsations
though. For example, in a much quoted, but unpublished 1968 preprint Castor
reduced the problem of turbulent convection to a set of 3 coupled
time-dependent diffusion equations for the three second order moments of the
vertical velocity fluctuations $w$ and the temperature fluctuations $\theta$,
namely $<\ngth ww\ngth >$, turbulent energy, $<\ngth w\theta\ngth >$ and
$<\ngth \theta\theta\ngth >$.  The set of equations was closed with a
down-gradient approximation for the respective fluxes, and with an expression
for the turbulent energy dissipation $\epsilon \propto e_t^\trhh$
(Eq.~\ref{ML}).  Kuhfu\ss (1986) also considers a 3-equation version of the 1D
recipe.

Recently, in a series of papers, Canuto (e.g. Canuto 1998) reexamined this
problem and extended this formalism to higher order in which the downgradient
approximations are avoided because they have been found lacking, in which
vertical-horizontal anisotropy is allowed for through the introduction of a
separate vertical turbulent energy, and finally, in which a dynamic equation
for the energy dissipation $\epsilon$ is introduced.  As a result there are now
five nonlinear coupled time-dependent diffusion equations to be solved together
with the hydrodynamics and radiation transport equations.  For the time being,
incorporating such a scheme into our pulsation code is a daunting task and we
prefer to get as much insight from a simplified one-equation treatment as
possible.

It is nevertheless of interest to see to which of the above recipes Canuto's
formalism reduces when additional, 'usual' simplifying assumptions are made.
Thus, starting with Canuto \& Dubikov's (1998, hereafter CD) \th
Eqs.~(CD19a--19d) we keep the first for $K$ ($\equiv e_t$) which is the
equivalent of our equation for the turbulent energy after we make the down
gradient approximation $D_f(e_t) \propto e_t^\ahh\nabla e_t$.  Instead of
Eq.~(CD19d) one can introduce an equation for the anisotropy turbulent energy
by subtracting 1/3$\times$ Eq.~(CD19a) from (CD19d).  In this equation and in
Eqs.~(CD19b--19c) one now makes a local time-independent approximation, \viz
${d/dt} \rightarrow 0 $, and ignores the diffusion terms $D_f$ and the $\chi$
terms (large P\'eclet number).  After some algebra this leads to

\vskip 6mm

 \begin{eqnarray}
    F_c     &= d_1 \th \rho\th c_p T\th\th 
             \bigl(e_t^2/\epsilon\bigr)\th\th  Y \th\th / {\cal D}
    &=  d_1 \th \rho\th c_p T\th e_t^\ah Y \th\th / {\cal D} \\
  {\cal S}  &= d_1 \th (\alpha_t /\Gamma_1) \th\th
                    \th \bigl(e_t^2/\epsilon\bigr) \th\th / {\cal D}
      &= d_1 \th (\alpha_t/ \Gamma_1)  \th\th\th 
                 c_s^2 \th e_t^\ah Y \th\th /   {\cal D}  \label{eq_canuto}\\
   & &  \nonumber \\
  {\cal D}   &= 1-d_2 \th  c_s^2 \th\th (e_t/\epsilon)^2\th\th
             &= 1 -d_2\th\alpha_\Lambda^2 \th \Bigl( c_s^2/ e_t \Bigr)\th Y  
 \end{eqnarray}

\vskip 3mm

\ni In each of these last three equations, the second expression is obtained
with the use of Eq.~\ref{ML}.  The $d_1$ and $d_2$ are positive dimensionless
constants of order unity.

Except for the appearance of a denominator ${\cal D}$, the CD formalism, with
the additional quoted approximations, thus leads to the Gehmeyr-Winkler recipe.
A quick inspection shows that the GW recipe is equivalent to ignoring
anisotropy, i.e. to setting $<\ngth w^2\ngth >=\twot \th e_t$, instead of using
Eq.~(CD19d), and to ignoring the $<\ngth \theta\theta\ngth >$ in Eq.\th
(CD19c).


 \begin{figure}
 \centerline{\psfig{figure=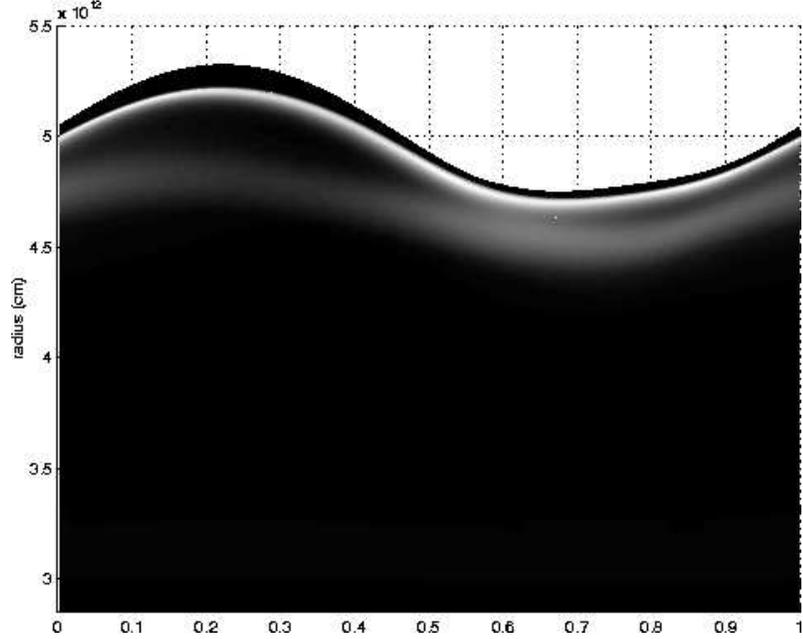,width=10.5cm}}
 \caption{\small 
 Variation of $e_t$ over a cycle as a function of radial coordinate $r$.  
 The lightness of the  shading indicates  the strength of $e_t$.
 } \label{fig_et2}
 \end{figure}


This denominator, however, has a pole because $d_2$ is positive and, if used in
a pulsation code, it would be disastrous.  Indeed, for small $e_t$,
Eq.~\ref{eq_canuto} implies that below a certain threshold any existing $e_t$
will decay away!  The reason for this unphysical denominator is either in the
starting formalism, or more likely, in the additional approximations that we
have introduced,


 \begin{figure}
 \centerline{\psfig{figure=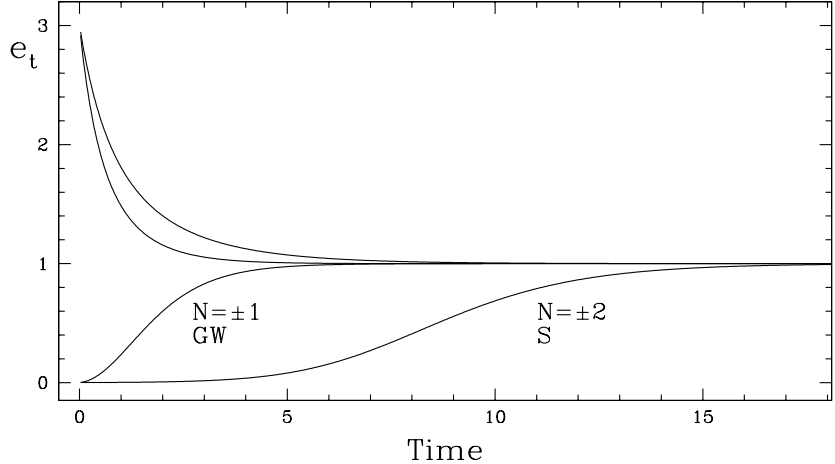,width=12.5cm}}
 \caption{\small 
  Integration of $de_t /dt = {\cal C} (e_t)$
  with small initial $e_t$ (bottom curves)
  and large initial $e_t$ (top curves) for the two values of $|N|$ 
  in Eq.~\ref{N}
 } \label{fig_tdep}
 \end{figure}


\begin{figure}
\centerline{\psfig{figure=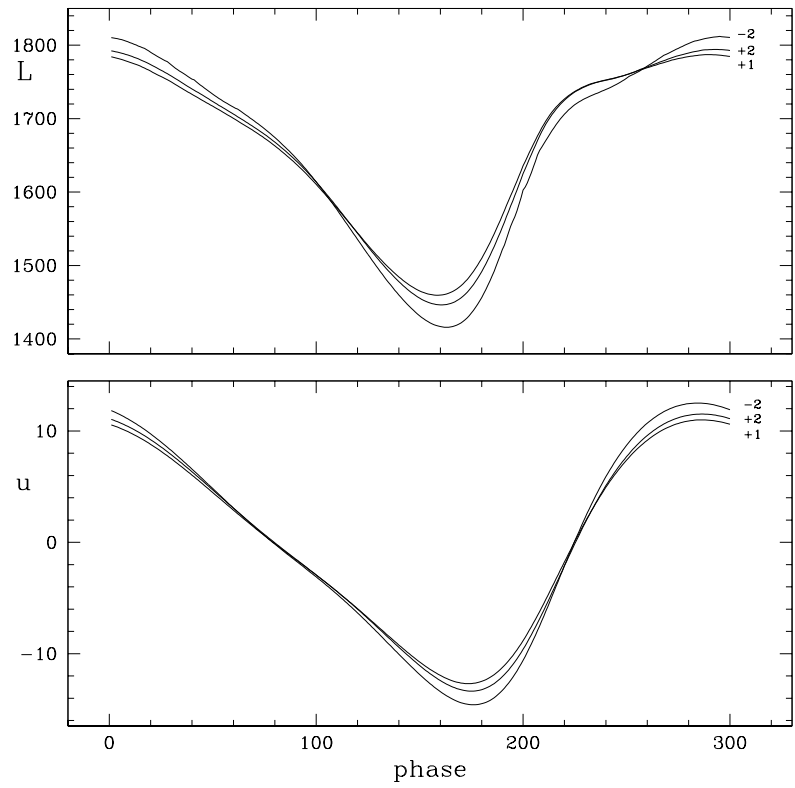,width=11.5cm}}
\caption{\small
Light curves ($L/L_o$) and radial velocity (km/s) curves for N=1 and $\pm$2 
} \label{fig_lc}
\end{figure}


The disregarded time derivatives of $<\ngth \theta\theta\ngth>$ and of
$J=<\ngth w\theta\ngth>$ ($\sim F_c$) is not justified when $e_t$ very small,
and can dominate when turbulence is about to set in.  To see this, we look at
the equations for small $e_t$ when we need to consider only the source terms in
Eqs.~(CD19) (the dissipation terms having a higher power of $e_t$) which leads
to e.g.

\begin{equation}
d^2 X/dt^2 = g \alpha \beta\th X = (\alpha T /\Gamma_1) {c_s^2\over
H_p^2} Y \th\th X \equiv 1/\tau_{gr}^2\th\th X
\end{equation}
where, to within factors of order unity, $X$ is any of the second order
moments, and $\tau_{gr}$ is the initial growth-time of $X$.  With this in mind
we can write $(1/X) (dX/dt) =(1/e_t) (d\th e_t/dt)= (1/e_t) g\alpha J$ (CD
Eq.~19a).  It is easy to see that when the time derivative dominates (at small
$e_t$) then, $<\ngth\theta\theta\ngth>$ instead of being proportional to $J$,
one becomes
 $$<\ngth\theta\theta\ngth> = 2 \beta (g \alpha) K$$
which thus removes the troublesome denominator.

It is possible to carry out the intended plan of only keeping Eq.~(CD 19a) and
of replacing the three equations (CD 19b--19d) by their local limit.  However,
as we have shown it is necessary to keep the derivative terms at the cost of a
higher order polynomial equation for $J$ ($F_c$) which prevents simple
analytical expressions such as Eq.~(\ref{eq_canuto}), and is not very
practical.

\section {Comparison of Recipes}

\subsection{Time-Dependence}

To give an indication how important time-dependence is in the pulsating Cepheid
envelope we show the behavior of the turbulent energy as a function of time
over a period for a typical cepheid model in Figure~\ref{fig_et2}.  The region
of maximum turbulent energy clearly undergoes substantial motion with respect
to the fluid.  It also shows a widening and intensification of the turbulent
region during the pulsational compression phase.

Because of this time-dependent behavior of the turbulent energy and because of
the differences in the expressions for the source terms in Eqs.~\ref{tc_rec} it
is of interest to see how $e_t$ responds in the various formulations.  Gehmeyr
and Winkler (1992) and Kuhfuss (1986) already compared some of the properties
of their scheme to that of Stellingwerf.  Here we now show some additional
comparisons of the various schemes.  In Figure~\ref{fig_tdep} we display the
temporal responses of $e_t$ to a constant entropy gradient $Y$. Thus we show
the results of the integration of

 \begin{equation}
 {d\th e_t\over dt} = {\cal C}
 \label{detdt}
\end{equation}
 for two initial values of $e_t$, a very small one, and a large one.
The temporal response is quite different for the two values of $|N|$, the
adjustment time for $e_t$ to a changing source $Y$ being much slower for the
larger $N$, i.e. for the Stellingwerf choice of source.  A linearization of
Eq.~\ref{detdt} shows that the kernel has a different $e_t$ dependence which
can lead to small differences in the spectrum of the eigenmodes.

\begin{figure}
\centerline{\psfig{figure=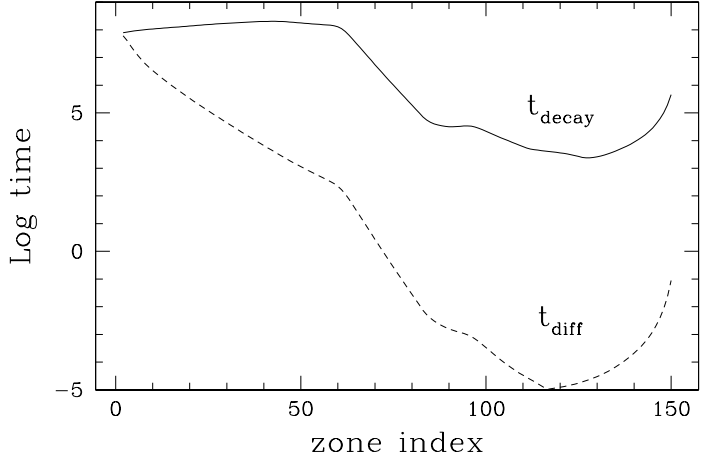,width=9.5cm}}
\caption{\small 
Variations of the turbulent diffusion and turbulent decay 
times (s) over the Cepheid
envelope.
} \label{fig_time}
\end{figure}


\subsection{Limit Cycles}

Because the different schemes have different timescales for the turbulent
energy, it is of interest to compare the nonlinear behavior of models.  In
Fig.~\ref{fig_lc} we show the fundamental limit cycles (light curves and radial
velocity curves) for a Cepheid model with $M$=5\Mo, $L$=1664\Lo, \Teff=5400\th
K, $X$=0.70, $Z$=0.02, ($\alpha_d$ = 4.0, $\alpha_c$ = 3.0, $\alpha_\Lambda$ =
0.35, $\alpha_s$ = 0.75, $\alpha_\nu$ = 1.5, $\alpha_t$ = 1, $\alpha_p$ =
0.67).  Results are displayed for the three formulations $N$=1 and $N=\pm$2.
The calculations actually show a great insensitivity to $N$, despite the
expectations of the previous section. The reason for this insensitivity is that
there are several time-scales at work, \viz a source time, a decay time and a
diffusion time for $e_t$.  In addition, for each of the time-scales varies by
many orders over the stellar envelope, as can be seen in Figure~\ref{fig_time}
which shows the run of the diffusion and of the decay time-scales for
the same Cepheid model.  The insensitivity is also in accord with the results
of YKB who showed that a 'sudden' approximation (${\cal C} = 0)$, i.e. an
instantaneous adjustment of $e_t$ to the local static equilibrium value gives
quite a good approximation to the exact linear eigenvalues.


\begin{figure}
\centerline{\psfig{figure=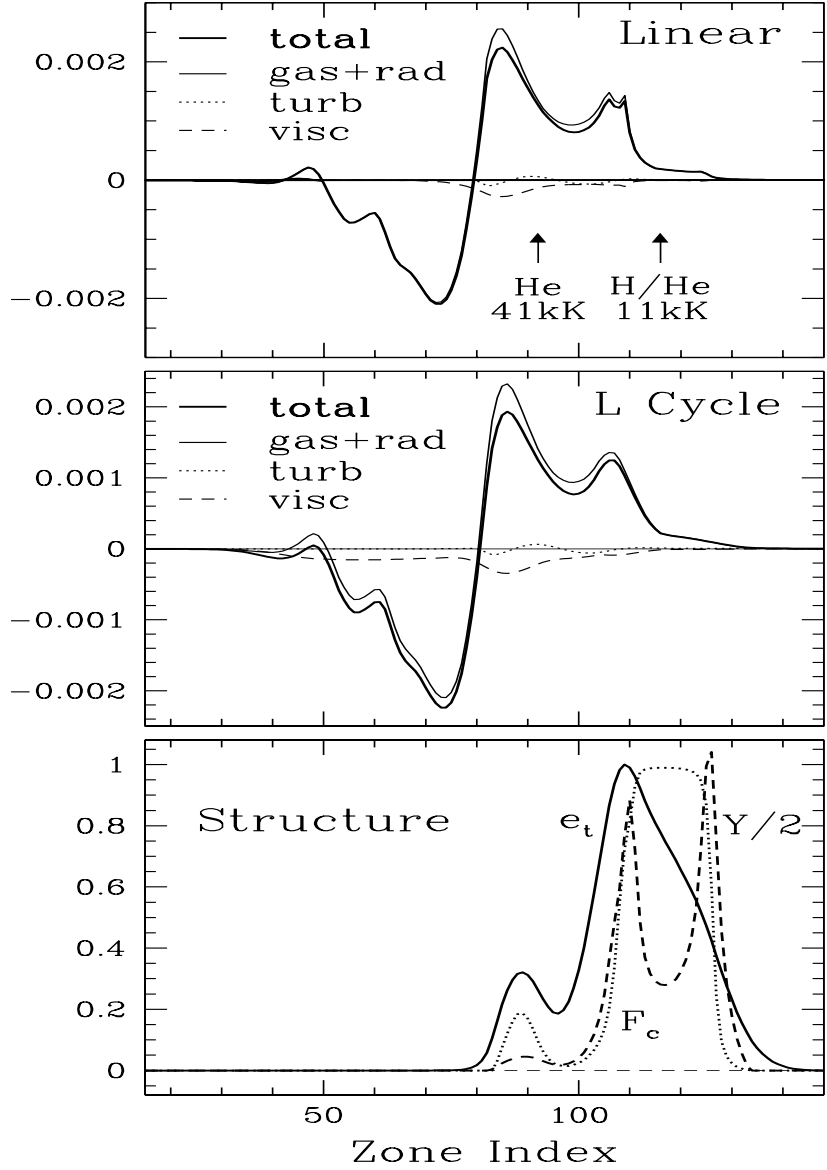,width=12cm,height=16truecm}}
\caption{\small 
Work integrands $<\ngth p dv\ngth >$ 
for a typical Cepheid model in the middle of the instability strip;
\th\th {\sl top}: linear; \th\th and {\sl middle}: nonlinear; \hfill\break
{\sl bottom}:  dimensionless entropy gradient $Y$, the turbulent energy, 
normalized to unity, and the fraction of
energy carried by the convective flux $F_c$.
}
 \label{fig_work}
\end{figure}


\subsection{Work Integrand}

It is well known that the pulsations of the classical variable stars are
self-excited through the $\kappa$ mechanism (e.g. Cox 1980, Gautschy \& Saio
1995).  When, for a given vibrational mode, the driving that occurs in the
partial ionization regions overwhelms the damping that occurs elsewhere, the
mode becomes unstable or self-excited.  This can best be demonstrated by the
work-integrands which show the cycle averages of $<\ngth p dv\ngth >$
throughout the star, i.e. the amount of internal energy converted into
mechanical (pulsational) energy.  The pressure that appears here is the total
pressure $p= p_{g+r} + p_t +p_\nu$.  We can thus look at the contributions of
these components separately.

In Figure \ref{fig_work} the total work-integrand for linear perturbations is
displayed as a thick line for the fundamental Cepheid model of the previous
section.  Shown with arrows are the points of half-ionization of H at 11,000\th
K and and He$^{I}$ 41,000\th K, respectively.  The molar tooth-shaped feature
on the right is produced by the combined H and first He ionization regions
which cause a broad convective region, with the spikes occurring at its
boundaries.  Just as in the purely radiative models the second ionization of He
produces both a driving and a damping region.

The turbulent pressure plays a very small role in general, although turbulent
pressure gradients can be important locally (YKB).  The eddy pressure, on the
contrary, is an important source of damping, especially in the limit cycle where
turbulent energy is spread over a wider range.  For reference, the bottom
figure shows the source of turbulence, namely the dimensionless entropy
gradient $Y$, the turbulent energy, normalized to unity, and the fraction of
energy carried by the convective flux $F_c$.  The driving is still positive in
the H/He partial ionization region even though $F_c$  is large there.
Some models, especially those with Z=0.02 can also have a convective region
associated with Fe-group elements.

As is typical in the partial H/He ionization region Cepheid envelopes,
convection reduces its source $Y$ by an order of magnitude, but it is not
capable of reducing it at the boundaries of the convective region to less than
a value of about 2--3.

\begin{figure}
\centerline{\psfig{figure=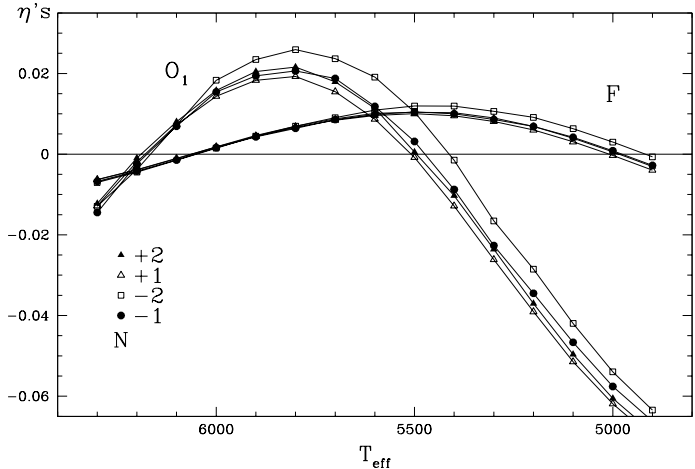,width=9.5cm}}
\caption{\small 
Behavior of the growth-rates ($\eta=2\kappa\times$ Period) of the fundamental
and of the first overtone modes as a function of \Teff\ for a sequence of
Cepheid models   ($M$=5\Mo, $L$=1664\Lo, $X$=0.70, $Z$=0.02)
} 
\label{fig_etas}
\end{figure}

\subsection{Sequence of Cepheid models}

In Fig.~\ref{fig_etas} we display the behavior of the relative growth-rates
$\eta$ of the fundamental and first overtone modes for a sequence of Cepheid
models ($M$=5\Mo, $L$=1664\Lo, $X$=0.70, $Z$=0.02). The $\eta$'s are defined as
$\eta=4\pi \kappa/\omega$ (where the linear eigenvalues are $\sigma = i\omega
+\kappa$) and represent the pulsational energy gain per cycle (the inverse of
the quality factor $Q$ in electronic circuits).  (The models for the four
sequences differ slightly because the small, but nonzero turbulent flux affects
the equilibrium.)  The four recipes are seen to give almost the same results,
again showing that the time-dependence of the recipes is not critical.  This is
also in agreement with what YKB  found, namely that the sudden approximation
(instant adjustment of $e_t$ to the instantaneous static solution) is an
excellent approximation.  This is in contrast to the frequently made, but bad
approximation of 'freezing' the convective flux in the computation of the
linear eigenvalues.

\vskip 10pt

We conclude this section by noting that the three recipes give very similar
behavior, and that the apparent differences in the published results are most
likely due to different choices of the $\alpha$ parameters than to the recipes
themselves.

\section{Results}

\subsection{Light-Curve Fourier Decomposition Coefficients}

In the Introduction we mentioned the presence of resonances in Cepheids between
the self-excited pulsational mode and an overtone, and that this is the reason
why the Fourier decomposition coefficients of the light-curves and of the
radial velocity curves show so much structure in the Cepheids.  RR~Lyrae are
devoid of such resonances and show a relatively dull behavior of the Fourier
decomposition coefficients as a function of period.  The presence of these
resonances puts very severe constraints on Cepheid models that any turbulent
convective description must satisfy (see below).

Purely radiative models were unable to reproduce the large excursion of the
``Z'' shape of the $\phi_{21}$ Fourier coefficient for the overtone Cepheids
(e.g. Buchler 1998, Antonello \& Aikawa 1993, 1995; Schaller \& Buchler, 1994,
unpublished preprint).  It came therefore as a pleasant surprise when the
nonlinear turbulent convective overtone Cepheid models showed the ability to
reproduce wide excursions in the $\phi_{21}$ such as indicated by the
observations.  However, since extensive nonlinear calculations are needed, a
systematic study of the Fourier coefficients remains to be made.

\subsection{Double-Mode Pulsations in Cepheids and RR Lyrae}

The numerical modelling of steady, nonlinear double mode (DM) pulsations has
been a long-standing quest in which purely radiative models have failed.  In a
recent paper (Koll\'ath \etal 1998) it was found that with the inclusion of a
1D turbulent convection recipe in the code,\th DM pulsations appear quite
naturally in Cepheid models.  Almost concomitantly, but independently,
Feuchtinger (1998b) encountered DM pulsations in RR Lyrae, which we have since
also confirmed.

Buchler \etal (1999) show that the behavior of the double mode phenomenon both
in Cepheids and in RR~Lyrae can be captured by rather simple amplitude
equations (also called normal forms).  Here $A_0$ and $A_1$ denote the modal
amplitudes of the two excited vibrational modes.
 \begin{eqnarray} 
 {d\th A_0\over dt}
  = \bigl (\kappa_0 -q_{00} A_0^2 -q_{01} A_1^2 \th -r_0 A_0^{\scriptstyle 4}
  \bigr )  \th\th A_0 \\ 
{d\th A_1\over dt}
  = \bigl (\kappa_1 -q_{10} A_0^2 -q_{11} A_1^2 \th -r_1 A_1^{\scriptstyle 4}
 \bigr ) \th\th A_1
 \end{eqnarray} 
The transient evolution of a hydrodynamics model is governed by these equations
(Koll\'ath \etal 1999).
Furthermore, and more importantly, the fixed points of these amplitude
equations ($dA_0/dt=dA_1/dt=0$) allow  an overview of the 'modal selection'
(bifurcation diagram) in the physical space of $L$ and \Teff, for example. 

\vskip 5pt

Koll\'ath \etal 1999 have computed double-mode pulsations with both the YKB and
the GW formulations (Eqs.~\ref{tc_rec}).  Both recipes give rise to DM
pulsations in broad domains of the physical ($L$, $M$,\Teff) and ($\alpha$)
parameter space.  We note that the DM behavior occurs in Galactic as well as in
MC Cepheid models (Koll\'ath \etal 1999).  

We have recently found that correcting the recipe (Eq.~\ref{eqs_tc} for small
P\'eclet number (see below) allows one to find also overtone double-mode
pulsations (as are observed), in which the first and second overtones are
excited.

We finish this section by stressing that the range of observed double-mode
periods, amplitude ratios and Fourier phases provides a powerful set of
constraints on the model parameters.  This is true not only for the DM
fundamental--1$^{st}$ overtone and 1$^{st}$--2$^{nd}$ pulsators, but also for
the single mode Cepheids and RR Lyrae.

\subsection{RR Lyrae pulsations}

 Extensive computations of convective RR~Lyrae model pulsations have been made
by Feuchtinger (1998a, 1999) and Feuchtinger \& Dorfi (1997) using the
Gehmeyr-Winkler recipe (Eqs.~\ref{tc_rec}), but omitting the turbulent flux and
the turbulent pressure, and by Bono \etal (1997), using the Stellingwerf
prescription.  These results are reviewed by Feuchtinger in this Volume.

\section{Observational Constraints}

In the Introduction we have pointed out that purely radiative models have
failed in many respects.  In the previous section we have discussed the
recent successes of the turbulent convective computations,
and we have compared them to the results of the prior radiative
modelling.  Despite the early optimism that these successes have generated,
there remain some very serious discrepancies.  That they are more apparent in
the Cepheid models than in RR~Lyrae is not astonishing.  RR~Lyrae stars,
despite their observed differences, are really very similar to each other in
mass, luminosity and composition.  The observational data of the Cepheids, on
the other hand, span more than an order of magnitude in luminosity and in mass,
and the metallicity of the Galactic Cepheids is four to five times that of the
SMC Cepheids.  These observations impose a number of important constraints that
we now address.

\subsection{Resonances and Mass--Luminosity relation}

In 1928 Hertzsprung noticed that a bump on the light curves of the fundamental
Cepheids that moved from the descending branch of the light curve to the
ascending one at a pulsational period $P_0\sim 10$d.  Later it was found that
this Hertzsprung bump progression and its corresponding manifestation in the
Fourier decomposition coefficients is related to a resonance of the fundamental
mode of pulsation with the second overtone (Simon \& Schmidt 1976, Buchler \&
Goupil 1984).
 
It is well established by now that structure in the behavior of the Fourier
decomposition coefficients of light curves and radial velocity curves as a
function of, usually, the period or effective temperature, is related to
resonances.  Because the structure of the stellar envelope changes with this
'control' parameter, the periods and period ratios also change and the excited
pulsation mode can run into a resonance condition with an overtone which then
gets entrained by this resonance through nonlinear coupling (e.g. Buchler
1993).  There are two reasons for the prominence of the Hertzsprung resonance.
First, the period ratio $P_0/P_2 = 2$ is small, making the modal coupling very
low order and thus strong.  Second, the second overtone, while linearly stable,
is only slightly stable, which allows it to be entrained to an appreciable
amplitude through the resonance.

For the first overtone Cepheids, there is also a 2:1 resonance which occurs in
the lightcurve data at a period around 3--4~d (Antonello \etal 1990).  More
recently this resonance has also been found to be quite prominent in the radial
velocity data (Kienzle \etal 1999).

The data from the microlensing surveys indicate that both these resonances
occur at the same pulsation periods, to within a day.  These constraints allow
us to compute 'resonance' masses and luminosities as a function of Z for stars
in the instability strip.  In other words it allows us to determine points on
the mass-luminosity curve that these stars must obey.  In a recent Letter
Buchler, Koll\'ath, Goupil \& Beaulieu (1996) discussed the implications of
these resonances for Cepheid models, and they showed that there is a serious
discrepancy between the pulsational models and stellar evolution calculations,
with in particular the pulsation masses being much too small for the SMC
models.  Those results were obtained with purely radiative models.  

We have since reexamined the problem with the turbulent convective models.  The
computed resonance luminosities still turn out largely independent of Z as
suggested by the observations (Beaulieu, private communication), and the
resonance masses are still much smaller for the lower Z values.  One notes
though that lower masses are in agreement with the evolutionary calculations
(e.g. Chiosi et al 1993, Schaller \etal 1992, Baraffe \etal 1998).  However,
reconciling these results with reasonable widths for the instability strips
remains a problem.

\subsection{Largest First Overtone Period}

The largest observed periods for the Galactic overtone Cepheids are $P_{1\th
max}=7\dotd 57$ for V440 Per, $P_{1\th max} =5\dotd 44$ for X Lac (Antonello
\etal 1990).  Beaulieu quotes $P_{1\th max} = 5\dotd 84$ for star No.~114 in
the LMC, $P_{1\th max}=5\dotd 98$ for star No.~255 in the SMC. The star V440
may be an oddball and if we can ignore it, then there appears to be a
metallicity independent upper limit $P_{1\th max}\approxgt 6$d.  The overtone
periods in the beat Cepheids do not provide any upper limit here, because all
of them are considerably smaller than the periods  of the single-mode overtone
Cepheids.

For the following discussion we refer to Fig.~12 of YKB that shows
Hertzsprung -Russell diagrams with the shapes of the fundamental and first
overtone instability strips (IS) for Cepheids.  The overtone IS is pinched off
above a certain luminosity (or period).  We refer to the overtone period at the
tip of the overtone IS as $P_{1\th max}$.

We can adjust the $\alpha$ parameters so that for Galactic Cepheid models
(typified by X=0.70, Z=0.02) we get an upper limit of $P_{1\th max}=6d$.  If,
with the same $\alpha$'s we now compute the $P_{1\th max}$ for SMC composition
(typified by X=0.726, Z=0.004) we obtain values exceeding 12~d for a broad
range of $\alpha$ parameters, which is clearly {\sl not} in agreement with the
observations.  Furthermore, if we take a larger $Z$ value for the Galaxy, as is
sometimes suggested, the discrepancy worsens.  Conversely, adjusting the
$\alpha$'s to get $P_{1\th max}$ in agreement with the SMC value gives maximum
overtone periods that are far too small for the Galaxy.

The problem arises because the lower masses and the low metallicity of the MC
Cepheids.  For the same luminosity and $\alpha$'s these stellar models are more
unstable (larger $\eta$'s) than their Galactic siblings, which pushes the
$P_{1\th max}$ to much higher values.  At this time a resolution of this
difficulty appears to pose a serious challenge to convective Cepheid modelling.

\subsection{Width of Instability Strip}

Adjusting our $\alpha$'s so that the Galactic $P_{1\th max}$ is $\sim 6\dotd 0$
we invariably obtain (at the luminosity for that overtone period) a width for
the fundamental IS that exceeds 1000K.  We consider such a width quite
excessive because in addition to the intrinsic width (for a given composition)
there must be substantial widening due to metallicity dispersion.

\begin{figure}
\centerline{\psfig{figure=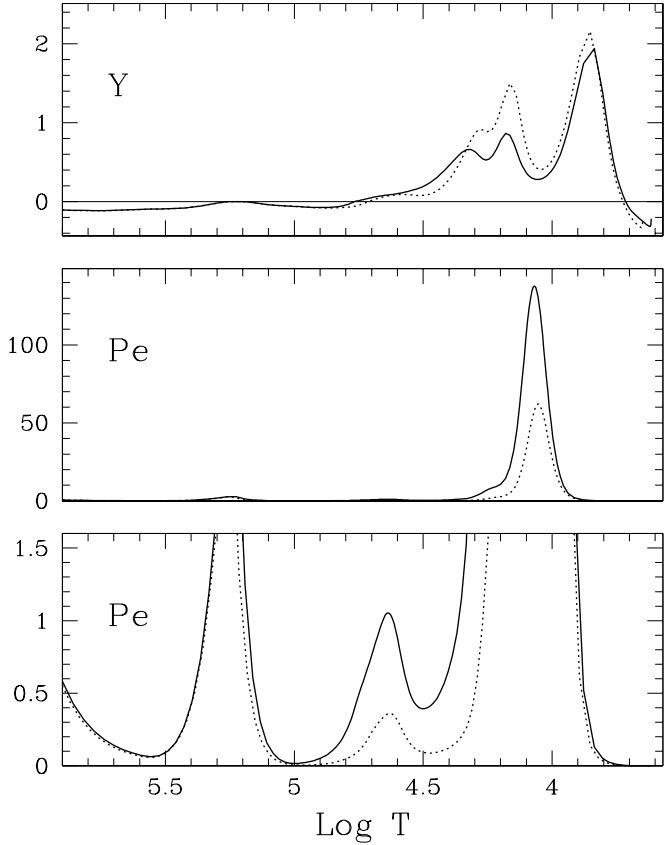,width=9.cm}}
\caption{\small
Top: Entropy gradient ($Y$);\th Middle: P\'eclet number, as a function of Log T
for Cepheid models: $M$=5\Mo, $L$=2000\Lo, X=0.70, Z=0.02; \Teff=4900\th K
(solid line), \Teff=5300\th K (dotted line);\th Bottom: Blow-up of small P\ngth
e number scale.
} \label{fig_peclet}
\end{figure}

We have found that this problem may have its origin in the formulation of the
convective recipes Eqs.~\ref{tc_rec}, that are all derived under the assumption
of efficient convection, i.e. of large P\'eclet number\th (P\ngth e
is the ratio of radiative to convective diffusion times, thus measures the
radiative damping of the convective elements).  But this assumption of large
P\ngth e is not satisfied everywhere in the Cepheid envelope.  In
Fig.~\ref{fig_peclet} we display the P\'eclet number throughout typical Cepheid
envelopes, namely $M$=5\Mo, $L$=5000\Lo, with \Teff=4600\th K (solid line) and
\Teff=5300\th K (dotted line).  As a guide we also show the behavior of the
dimensionless entropy gradient $Y$ in the upper figure.  The P\'eclet number is
very large throughout most of the combined H/He$^{\rm I}$ partial ionization
region, but as the blown up bottom picture shows, the second helium ionization
region lies entirely in the low P\ngth e regime.  There is also a small surface
region, below T=8000\th K, where P\ngth e $<$ 1.  Since the second ionization
stage of helium contributes substantially to the work-integrand we expect that
an improved treatment of ineffective convection has a major effect on the
growth-rates. 

In order to accommodate the small P\ngth e limit CD suggest an interpolation
(see also Kuhfu\ss\th 1986), which after a small manipulation can be seen to be
equivalent in the GW scheme to limiting $J$, i.e. both $F_c$ and ${\cal S}$ in
Eqs.~(\ref{tc_rec}) by a factor of the form

 $$f_{pec} = {1\over 1+\alpha_r P\ngth e^{-1}}\quad ,$$
where the P\'eclet number is the ratio of convective and radiative the 
diffusion coefficients,  P\ngth e$=  D_c/D_r $, with 
 $$D_r = {4\over3} {a c T^3 \over \kappa\rho^2 c_p}$$
 and 
 $$D_c = \Lambda e_t^\ahh$$
  
\vskip 5pt

The definition of P\ngth e is somewhat arbitrary and we have introduced an
additional free, order unity, parameter $\alpha_r$.  Both the convective flux
and the source term thus become proportional to P\ngth e when P\ngth e is
small, and scale with an additional factor of $e_t^\ahh$.  As a result, in the
static equilibrium model (if we disregard diffusion), one obtains $e_t \sim
Y^2$, instead of $e_t \sim Y$, and both $F_c$ and ${\cal S}$ scale with
$Y^3$, instead of $Y^\trhh$.

Referring back to Fig.~\ref{fig_peclet}, we note that in the higher mass
Cepheid models, such as this one, a strong ($Y$ very small) convective zone can
appear in the Fe ionization region $\sim$200,000\th K.

\begin{figure}
\centerline{\psfig{figure=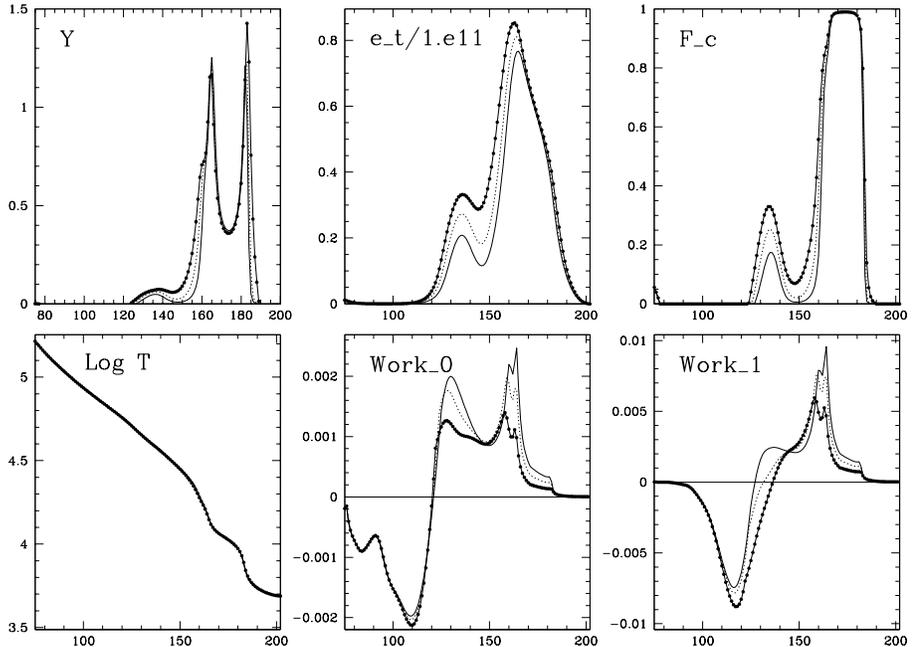,width=12.cm}}
\caption{\small
Effect of P\'eclet correction factor on, from left to right, top down, 
as a function of zone index, the entropy
gradient Y, the turbulent energy $e_t$, the convective flux ($F_c/F_{tot}$),
the temperature versus zone index, the turbulent flux $F_t/F_{tot}$, the
work-integrands of the fundamental and first overtone modes. \hfill\break
Solid-dotted line: $\alpha_r$=0, dotted lines: $\alpha_r$=0.1 
and thin line: $\alpha_r$=0.25. 
}
 \label{fig_alphar}
\end{figure}

We have performed some preliminary calculations with the P\'eclet factor
$f_{pec}$.  Figure~\ref{fig_alphar} shows the change of some of the model
properties for $M$=5\Mo, $L$=3184, $X$=0.70, $Z$=0.02, with $\alpha_r$=0, 0.1
and 0.25.  ($\alpha_c$ = 4.0, $\alpha_\nu$ = 1.0, $\alpha_s$ = 0.75, $\alpha_p$
= 0.667, $\alpha_t$ = 1.0, $\alpha_d$ = 4.0, $\alpha_\Lambda$ = 0.58).  One
sees the reduced efficiency of turbulent energy source particularly in the H/He
region.  This figure is somewhat misleading though.  The P\'eclet factor
decreases the turbulent energy which means that in order to satisfy the
observational constraints, we have to compensate by increasing some of the
$\alpha$'s, which is not reflected in the figure.

However, the important result is that the introduction of $f_{pec}$ has a
differential effect on the stability of the fundamental and first overtone
modes that goes in the direction of solving the problem of the excessive width
of the fundamental instability strip.  That width is most sensitive to the
parameter $\alpha_r$ which can thus quite satisfactorily be used for adjusting
the width to satisfy this observational constraint.


\subsection{RR Lyrae}

For RR Lyrae the 'linear' constraints are somewhat different.  Here, because
globular clusters have approximately the same luminosity, we know the edges of
the corresponding instability strips directly from the observed fundamental
(RRab) and first overtone (RRc) periods.  The observed \Teff\ are probably too
uncertain to be directly very useful, although for specific globular clusters
stars they may provide useful constraints on the temperature widths of the ISs.
However, it is important to keep in mind that using the linear
periods for the constraints can be misleading because a model may be inside the
linear instability strip, yet the corresponding limit cycle may be unstable
(e.g. Buchler \& Kov\'acs 1986).  In other words, the observational instability
strip is narrower than the linear instability strip, and it takes nonlinear
calculations to determine the former.

A great deal of observational lightcurve data are available, and their Fourier
decomposition coefficients provide strong nonlinear constraints that need to be
satisfied by the models.  In particular there is a nice correlation between the
coefficients that has been discussed by Kov\'acs and Kanbur (1997).

The double-mode, RRd stars also provide tight constraints on the alpha
parameters through the observed
periods and period ratios, and through the amplitude ratios.


\subsection{Additional Constraints}

From the observational data one can infer period -- radius relations for
Cepheids (e.g. Laney \& Stobie 1994).  These constraints are very basic since
they concern the equilibrium models.  Bono \etal (1998, 1999) have 
had mixed success reproducing this observational constraint.

Other nonlinear interesting constraints arise from the temperature fluctuations
of the Cepheids and RR Lyrae.  Simon, Kanbur \& Mihalas (1993) have shown that
essentially all Cepheids have the same $T_{max}$ (same spectral type), but that
$T_{min}$ depends on the amplitude of pulsation. Similarly, but oppositely, RR
Lyrae have essentially the same $T_{min}$ (e.g. Kanbur \& Phillips 1996).
These constraints have not yet been taken into account in the calibration of
the $\alpha$ parameters.

\section{Theoretical Difficulties}

For a given set of $\alpha$s the 1D recipes all give the same turbulent energy
$e_t$ in the equilibrium models when the turbulent flux is neglected, which is

\begin{equation}
e_t = \alpha_s^2 \th\alpha_\Lambda^2 \th {\beta T\over \Gamma_1} \th c_s^2
\th Y {\rm \quad or \quad} u_t \sim c_s \sqrt{Y} 
\label{eq_et}
\end{equation}
 When this $e_t$ is inserted in the expression for convective flux
one obtains

\begin{equation}
F_c = \alpha_c\alpha_\Lambda \th (\rho c_p T) \th c_s Y^\trh
\label{eq_Fc}
\end{equation}
 However, $F_c$ is physically limited to transporting the available energy
fluctuations whose very generous upper limit is $(\rho c_p T)$.  Furthermore,
under the underlying assumptions the convective velocity should be
subsonic. Eq.~(\ref{eq_Fc}) clearly exceeds this upper limit when $Y$ is larger
that unity (cf. Fig.~\ref{fig_peclet}).  For this reason ad hoc flux limiters
have been proposed (e.g. Feuchtinger 1999)

The ultimate reason for the breakdown of the 1D recipe is twofold.  First, in
the convective boundary regions where $Y\approxgt 1$, the turbulent Mach number
$M_t=(u_t/c_s)$ invariably becomes important and can even exceed unity in some
weakly convective models.  However, it is well known that the approximations
leading to the recipes Eqs.~(\ref{eqs_tc}) break down, in particular because
pressure fluctuations were neglected in their derivations and the dissipation
term $\epsilon$ does not take into account larger Mach numbers.  Second, as
stressed by Canuto (1998) the downgradient approximations for the fluxes (here
$F_c \sim ds/dr$) are not very satisfactory.  Neither of these problems can
easily be corrected within a practically useful 1D recipe, and a flux limiter
may be the most expedient available patch.

\section{Conclusions}

The introduction of a simple recipe for turbulent convection, namely a 1D
time-dependent diffusion equation for the turbulent energy and the concomitant
expressions for convective and turbulent fluxes, eddy viscous pressure and
turbulent pressure, have provided a substantial improvement over purely
radiative models.  Perhaps the most remarkable achievement is the successful
modelling of double-mode pulsations both in Cepheids and in RR Lyrae.
Correcting the recipes for low P\'eclet number provides a further definite
improvement.

However, a number of problems persist as we have seen.  The number (8) of free,
order unity parameters is large which makes a search for optimal values quite
burdensome.  We still hope that it will be possible to find a range of suitable
parameters that satisfy all the observational constraints, and this
independently of metallicity or of the stellar type.  If not we will be forced
to adopt more complicated formulations of convective transport.

\acknowledgments

We are much obliged to Phil Yecko and Michael Feuchtinger for many discussions.
Phil kindly provided us with Figure~1.  We also wish to thank Vittorio Canuto
for fruitful conversations and for suggesting the use of his formalism to
discriminate among the various one-equation approximations.  Finally, we
gratefully acknowledge the support of NSF (AST95--28338) and ZK the support 
of OTKA (T-026031).

 \end{document}